\documentclass[aps,pr,superscriptaddress,preprintnumbers,nofootinbib,10pt]{revtex4-2}

\usepackage{multirow}
\usepackage{amsmath}
\usepackage{amssymb}
\usepackage[dvipdf,dvips]{graphicx}
\usepackage{color}
\usepackage{hyperref}
\usepackage{url}
\usepackage{slashed}
\usepackage{subfigure}
\usepackage[usenames,dvipsnames]{xcolor}
\usepackage{amsmath}
\usepackage{amsfonts}
\usepackage{float} 
\usepackage{amssymb}
\usepackage{epsfig}
\usepackage{graphics}
\usepackage{euscript}
\usepackage{slashed}
\usepackage{epstopdf}
\usepackage[utf8]{inputenc}
\allowdisplaybreaks
\usepackage[normalem]{ulem}
\usepackage{pifont}
\usepackage{dsfont}
\usepackage{MnSymbol}
\usepackage{verbatim}
\usepackage{graphicx}
\usepackage{latexsym}
\usepackage{tikz-feynman}
\usepackage{bbold}

\begin{document}

	\title{{\bf Entangled Schr{\"o}dinger cat states, vacuum projector  and Bell-CHSH inequality}}

	\author{S. P. Sorella} \email{silvio.sorella@fis.uerj.br} \affiliation{UERJ $–$ Universidade do Estado do Rio de Janeiro,	Instituto de Física $–$ Departamento de Física Teórica $–$ Rua São Francisco Xavier 524, 20550-013, Maracanã, Rio de Janeiro, Brazil}

		\begin{abstract}
Use of the vacuum projector and of the unitary displacement operators enables us to construct Hermitian dichotomic operators. These operators are employed to scrutinize the violation of the Bell-CHSH inequality for entangled coherent  Schr{\"o}dinger cat states. 
	\end{abstract}

	\maketitle

	\section{Introduction}\label{Intr}

	The entanglement of coherent states is a fascinating topic, both from theoretical as well as from experimental points of view \cite{Sanders:2012pme}. \\\\Depending on the size of the mean value of the occupation numbers, these states can be employed to account for regimes   ranging from quantum to semi-classical ones. \\\\In this note we devise the construction, out of the vacuum projector $|0 \rangle \langle 0| $, of  Hermitian dichotomic operators which can be used to study the violation of the Bell-CHSH inequality\footnote{See \cite{Guimaraes:2024byw} for a general  introduction to the subject.} 
\cite{Bell64,CHSH69}. \\\\As for the quantum state, an entangled Schr{\"o}dinger cat type state will be considered, leading to expressive violations of the Bell-CHSH inequality. \\\\The present framework  displays several advantages with respect to previous proposals \cite{JQ,Sorella:2024rwx} as, for example:
\begin{itemize}
\item the whole construction relies fully on the use of the vacuum projector $|0 \rangle \langle 0| $ and of the displacement operators, resulting in simple  Hermitian dichotomic operators. 

\item the Bell-CHSH correlation function can be easily evaluated in a closed compact  analytic form, a feature very helpful for the analysis of the corresponding violation

\item in principle, the Hermitian dichotomic character of the aforementioned operators might lead to possible experimental setups for further studies of the entanglement properties of cat type states. Needless to say, the devices available nowadays in Quantum Optics allow for nice experimental handlings of coherent states, see \cite{Sanders:2012pme} and refs. therein.

\end{itemize}

The paper is organized as follows. In Sect.\eqref{vp} we present the construction of  Hermitian dichotomic operators out of the vacuum projector. Sect.\eqref{BCHSH} is devoted to the analysis of the violation of the Bell-CHSH inequality, In Sect.\eqref{Conclusion}  we collect the conclusion. 
	
\section{Construction of Hermitian dichotomic operators out of the vacuum projector}\label{vp}	

Let ${\cal H} = {\cal H}_A \otimes {\cal H}_B$ be the Hilbert space of two harmonic oscillators $(a,b)$ 
\begin{eqnarray} 
[ a, a^{\dagger}] & = & 1 \;, \qquad [a,a]=0 \;, \qquad [a^{\dagger}, a^{\dagger}] = 0 \;, \nonumber \\
 \left[ b, b^{\dagger} \right] & = & 1 \;, \qquad [b,b]=0 \;, \qquad [b^{\dagger}, b^{\dagger}] = 0 \;, \nonumber \\ 
  \left[a,b \right] & = & 0 \;, \qquad [a, b^{\dagger}] =0 \;.  \label{ABab}
\end{eqnarray}
A coherent state $|\xi\rangle_{a}$ is obtained upon acting with the displacement operator ${\cal D}_a(\xi)$ on the vacuum state, namely 
\begin{equation}  
 |\xi\rangle_{a} = {\cal D}_a(\xi) |0\rangle =\; e^{(\xi a^{\dagger} - \xi^{*} a)} |0\rangle = \; e^{-\frac{|\xi|^2}{2}} e^{\xi a^{\dagger}} e^{-\xi^{*} a}|0\rangle \;, \label{coh}
\end{equation}
where $\xi$ is a complex number. The displacement operator ${\cal D}_a(\xi)$ is unitary, fulfilling the so-called Weyl algebra
\begin{equation} 
{\cal D}_a(\xi) \;{\cal D}_a(\xi)^{\dagger} = 1 = {\cal D}_a(\xi)^{\dagger} \;{\cal D}_a(\xi) \;, \qquad {\cal D}_a(\xi)\;{\cal D}_a(\xi') = e^{i(\Im(\xi) \Re(\xi') - \Re(\xi) \Im(\xi'))} \; {\cal D}_a(\xi+\xi') \;. \label{wa}
\end{equation}
From these relations, one gets 
\begin{equation}
\langle \xi \;|\; \xi \rangle_a = 1 \;, \qquad \langle \xi \;|\; -\xi \rangle_a = e^{-2 |\xi|^2} \;. \label{nm}
\end{equation}
One sees that, for suitable large values of $|\xi|$, {\it i.e.} $|\xi|>1$, 
\begin{equation} 
\langle \xi \;|\; -\xi \rangle_a \simeq  0 \;, \qquad |\xi| \;\;\; {\rm large} \;. \label{lg} 
\end{equation}
Already for $|\xi| = 2$ one has 
\begin{equation} 
\langle \xi \;|\; -\xi \rangle_a\Big|_{|\xi|=2} \simeq  0.00033
\end{equation}
In this regime, the two states $(|\xi\rangle_a, |-\xi\rangle_a)$ can be thought as normalized orthogonal states, much alike the  spin $1/2$ states $(|+>,|->)$. The superposition 
\begin{equation} 
|\psi\rangle_{\xi} = \frac{|\xi\rangle_a + |-\xi\rangle_a}{\sqrt{2}} \label{sc}
\end{equation} 
is usually referred to as a Schroedinger cat state. \\\\Let us move now to the bipartite Hilbert space ${\cal H} = {\cal H}_A \otimes {\cal H}_B$. In order to construct Hermitian dichotomic operators, we consider the vacuum projector 
\begin{equation} 
|0\rangle \langle 0| \;, \qquad |0\rangle = |0\rangle_a \otimes |0 \rangle_b   \;. \label{vproj}
\end{equation}
We can thus introduce the dichotomic operator 
\begin{equation} 
{\cal F} = 1 - 2 |0\rangle \langle 0| \;, \qquad {\cal F}^2 = 1 \;. \label{dicF}
\end{equation} 
Further, we act with the displacement operators $({\cal D}_a, {\cal D}_b)$ to define the four operators 
\begin{eqnarray} 
A(z) & = & {\cal D}_a^{\dagger}(z)\; {\cal F}\; {\cal D}_a(z) \;, \qquad A(z')= {\cal D}_a^{\dagger}(z')\; {\cal F}\; {\cal D}_a(z') \;, \nonumber \\
B(z) & = & {\cal D}_b^{\dagger}(w) \;{\cal F}\; {\cal D}_b(w) \;, \qquad B(w')= {\cal D}_b^{\dagger}(w')\; {\cal F}\; {\cal D}_b(w') \;, \label{four}
\end{eqnarray} 
with $(z,z',w,w')$ complex numbers.\\\\These operators obey the required conditions to be employed in the study of the Bell-CHSH inequality \cite{Guimaraes:2024byw}:
\begin{eqnarray} 
A(z) &  =  & A^{\dagger}(z) \;, \qquad A^2(z)=1 \;, \qquad A(z')   =   A^{\dagger}(z') \;, \qquad A^2(z')=1 \;, \nonumber \\
B(w) &  =  & B^{\dagger}(w) \;, \qquad B^2(w)=1 \;, \qquad B(w')   =   B^{\dagger}(w') \;, \qquad B^2(w')=1 \;, \label{cond1}
\end{eqnarray}
and 
\begin{eqnarray} 
\left[ A(z), B(w) \right] & = & 0 \;, \qquad \left[A(z), B(w')\right] =0\;, \qquad \left[ A(z'), B(w) \right]  =  0 \;, \qquad \left[A(z'), B(w')\right] =0\;, \nonumber \\
\left[ A(z), A(z') \right] & \neq 0 & \;, \qquad  \left[ B(w), B(w') \right] \neq 0 \;. \label{cond2}
\end{eqnarray} 

\section{The violation of the Bell-CHSH inequality}\label{BCHSH}

\begin{figure}[t!]
	\begin{minipage}[b]{0.4\linewidth}
		\includegraphics[width=\textwidth]{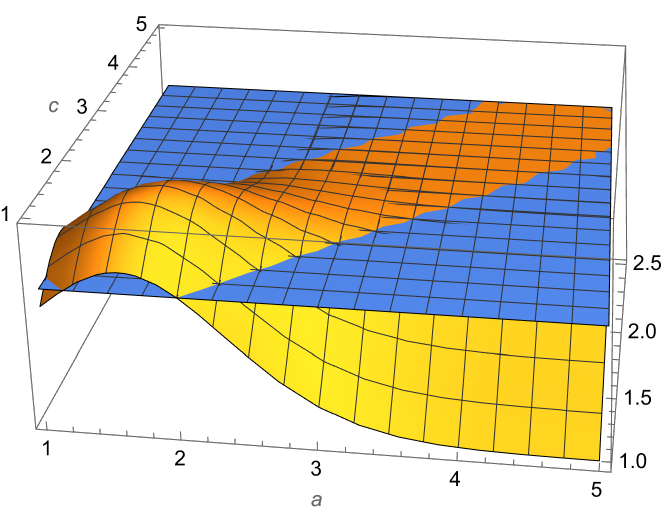}
	\end{minipage} \hfill
\caption{Behavior of the Bell-CHSH correlator  $\langle {\cal C} \rangle$ as a function of the parameters $(\alpha,\omega)$. The violation region corresponds to the region in which the orange surface lies above the blue one, which corresponds to the classical bound 2.}
	\label{Fig1}
	\end{figure}

\begin{figure}[t!]
	\begin{minipage}[b]{0.4\linewidth}
		\includegraphics[width=\textwidth]{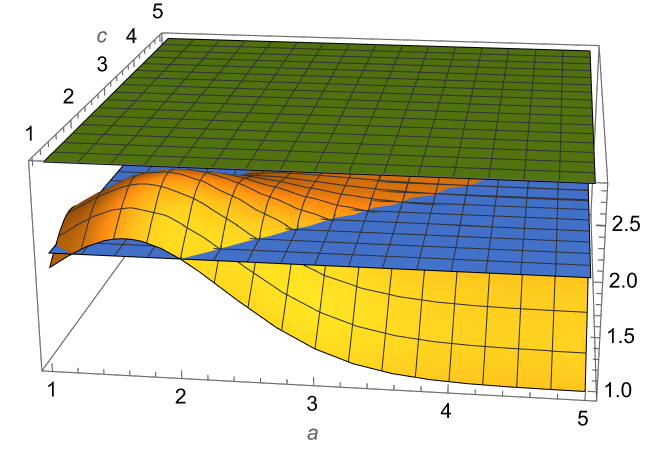}
	\end{minipage} \hfill
\caption{Behavior of the Bell-CHSH correlator  $\langle {\cal C} \rangle$ as a function of the parameters $(\alpha,\omega)$. The violation region corresponds to the region in which the orange surface lies above the blue one, which corresponds to the classical bound 2. The green surface is the Tsirelson bound $2\sqrt{2}$.}
	\label{Fig2}
	\end{figure}

We are ready now to scrutinize the Bell-CHSH inequality 
\begin{equation} 
\langle \psi | \; {\cal C} \; |\psi \rangle = \langle \psi |\; (A(z) + A(z'))\otimes B(w) + (A(z) - A(z'))\otimes B(w') \; |\psi \rangle  \;. \label{bchsh}
\end{equation}
A violation tales place whenever 
\begin{equation} 
2<  \Big| \; \langle \psi | \; {\cal C} \; |\psi \rangle \; \Big| \le 2 \sqrt{2} \;. \label{vt}
\end{equation}
The value $2 \sqrt{2}$ is known as the Tsirelson bound \cite{TSI}. We shall make use of a state $|\psi\rangle$ made up by two entangled cat type states, namely 
\begin{equation} 
|\psi \rangle = {\cal N} \left( {\cal D}_a(\sigma) {\cal D}_b(\eta) + e^{i\varphi}\; {\cal D}_a(-\sigma) {\cal D}_b(-\eta) \right) |0\rangle \;, \label{stp}
\end{equation}
where $(\sigma, \eta)$ are complex numbers, $\varphi$ is a phase factor  and the normalization factor ${\cal N}$ is given by 
\begin{equation} 
{\cal N} = \frac{1}{\sqrt{2}} \frac{1}{(1+\cos(\varphi) \;e^{-2(|\sigma|^2 + |\eta|^2)})^{1/2}} \;. \label{nf}
\end{equation} 
The Bell-CHSH correlation function, eq.\eqref{bchsh}, is easily obtained by making use of the following expression
\begin{eqnarray} 
\langle \psi|\; A(z) &\otimes&  B(w) \; |\psi  \rangle  =  1 + 4{\cal N}^2 \left( e^{-\frac{1}{2}( |z-\sigma|^2 + |w-\eta|^2)} + e^{-\frac{1}{2}( |z+\sigma|^2 + |w+\eta|^2)}  \right)  \nonumber \\
&+ &8 {\cal N}^2 \left( \cos( \varphi + 2 (\Im(z) \Re(\sigma) - \Re(z) \Im(\sigma) + \Im(w)\Re(\eta) - \Re(w) \Im(\eta) )) \right)e^{-\frac{1}{2}( |z-\sigma|^2 + |w-\eta|^2 + |z+\sigma|^2+ |w+\eta|^2)} \nonumber \\
& - &2 {\cal N}^2 e^{-|\eta|^2} \left( e^{-|z-\sigma |^2 }  +   e^{-|z+\sigma|^2 } + 2 \cos( \varphi + 2 (\Im(z) \Re(\sigma) - \Re(z) \Im(\sigma) ) )  e^{-\frac{1}{2}( |z-\sigma |^2 +|z+\sigma|^2}\right) \nonumber \\
& - &2 {\cal N}^2 e^{-|\sigma|^2} \left( e^{-|w-\eta |^2 }  +   e^{-|w+\eta|^2 } + 2 \cos( \varphi + 2 (\Im(w) \Re(\eta) - \Re(w) \Im(\eta) ) )  e^{-\frac{1}{2}( |w-\eta |^2 +|w+\eta|^2}\right) \;. \label{ABpsi}
\end{eqnarray}
In order to capture the violation of the Bell-CHSH inequality we set 
\begin{equation} 
\sigma = \alpha + i 0 \;, \qquad \eta = \omega + i 0 \;, z=z'=w=w'=1 \;, \qquad \varphi = \pi 
\end{equation}
and plot $\langle {\cal C} \rangle$ as a function of the parameters $(\alpha,\omega)$. The output is shown in Figures \eqref{Fig1} and \eqref{Fig2}. The blue surface is the classical bound 2, while the green one stands for the Tsirelson bound $2\sqrt{2}$. The orange surface shows the behavior of 
 $\langle {\cal C} \rangle$ as a functions of $(\alpha,\omega)$. One notices a rather big region in which the orange surface is above the blue one, displaying an expressive violation of the Bell-CHSH inequality.

\section{Conclusion}\label{Conclusion}

In the present work, we have presented a simple and elegant construction of Hermitian dichotomic operators useful for the study of the Bell-CHSH inequality for Schr{\"o}dinger cat type states. \\\\The whole setup relies on the use of the vacuum projector $|0 \rangle \langle 0|$  and of the displacement operators, as displayed by expressions \eqref{four}. The resulting operators give rise to expressive violations of the Bell-CHSH inequality, as reported in Figures \eqref{Fig1} and \eqref{Fig2}. \\\\The generalization to Mermin \cite{Mermin} inequalities is straightforward.

\section*{Acknowledgments}
	The author would like to thank  D. Azevedo, , F. Guedes, M.S. Guimaraes  I. Roditi and A. F. Vieira for discussions and collaborations. The Brazilian agencies CNPq and FAPERJ  are gratefully acknowledged  for financial support.  S.P.~Sorella is a CNPq researcher under contract 301030/2019-7.

\end{document}